# PRESERVING PHONEMIC DISTINCTIONS FOR ORDINAL REGRESSION: A NOVEL LOSS FUNCTION FOR AUTOMATIC PRONUNCIATION ASSESSMENT

*Bi-Cheng Yan[1], Hsin-Wei Wang[1], Yi-Cheng Wang[1], Jiun-Ting Li[1], Chi-Han Lin[2], Berlin Chen[1]*

National Taiwan Normal University, Taipei, Taiwan
E.SUN Financial Holding Co., Ltd., Taipei, Taiwan

{ bicheng, hsinweiwang, yichengwang, 60947036s, berlin}@ntnu.edu.tw; finalspaceman-19590@esunbank.com

## ABSTRACT

Automatic pronunciation assessment (APA) manages to quantify the pronunciation proficiency of a second language (L2) learner in a language. Prevailing approaches to APA normally leverage neural models trained with a regression loss function, such as the mean-squared error (MSE) loss, for proficiency level prediction. Despite most regression models can effectively capture the ordinality of proficiency levels in the feature space, they are confronted with a primary obstacle that different phoneme categories with the same proficiency level are inevitably forced to be close to each other, retaining less phoneme-discriminative information. On account of this, we devise a phonemic contrast ordinal (PCO) loss for training regression-based APA models, which aims to preserve better phonemic distinctions between phoneme categories meanwhile considering ordinal relationships of the regression target output. Specifically, we introduce a phoneme-distinct regularizer into the MSE loss, which encourages feature representations of different phoneme categories to be far apart while simultaneously pulling closer the representations belonging to the same phoneme category by means of weighted distances. An extensive set of experiments carried out on the speechocean762 benchmark dataset demonstrate the feasibility and effectiveness of our model in relation to some existing state-of-the-art models.

***Index Terms***— Automatic pronunciation assessment, computer-assisted pronunciation training, deep regression models, ordinal regression models

## 1. INTRODUCTION

Computer-assisted pronunciation training (CAPT) systems have become increasingly popular and been used for a multitude of use cases on language learning, with the purpose to enable learners to practice their speaking skills, alleviate the workloads of teachers [1], and others [2][3][4]. CAPT research can hark back to the middle of the last century [5] and has aroused increasing attention in recent years, showing impressive performance by leveraging many advanced machine learning technologies [6][7][8]. In common CAPT systems, second language (L2) learners are initially presented with a text prompt and instructed to read it aloud. By working in conjunction with the input speech and the presented text prompt, CAPT systems can access the learner's speaking proficiency and immediately provide instructive diagnostic feedback [9][10][11]. Through persistent repetition and practice, it is anticipated that L2 learners can gradually improve their speaking skills.

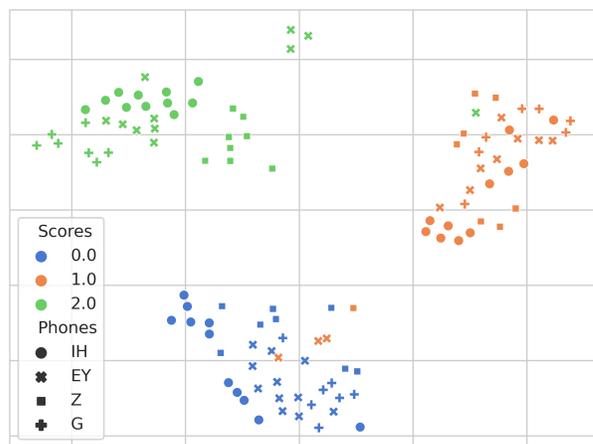

**Fig. 1**. Visualization of phoneme representations for a regression-based APA (GOPT) model trained with the mean-squared error loss reveals that the regression model tends to cluster features according to their proficiency scores.

According to the types of diagnostic feedback, the research endeavors of CAPT fall into two broad categories: one is phoneme-level mispronunciation detection and diagnosis (MDD), and the other is automatic pronunciation assessment (APA). The former aims to pinpoint phoneme-level erroneous pronunciations and provide L2 learners the corresponding diagnostic feedback [12][13][14]. The latter, in contrast, concentrates more on assessing and providing cross-level pronunciation scores to reflect the learner's pronunciation quality on some specific aspects or traits of their spoken language usage [15][16][17]. To this end, it evaluates pronunciation proficiency at various linguistic granularities (i.e., phoneme, word and utterance), with diverse aspects (e.g., accuracy, fluency and completeness). Contrary to MDD systems, APA systems possess the capability to provide a broad spectrum of pronunciation scores, making them conducive for L2 learners to improve their speaking skills more comprehensively. Consequently, APA systems have received much attention form academic and commercial sectors, yet remains a challenging task [18][19]. Prevailing approaches to the APA task usually follow a supervised training paradigm, where various neural models are employed to predict continuous scores that mimics human experts' evaluations on learners' speaking proficiency. Attributed to the continuous nature of the target output, these models are typically optimized with a regression loss function, namely mean-squared error (MSE). The MSE loss effectively preserves the inherent ordinality of the target output in the feature space by minimizing the average squared difference between the model predictions and the human evaluations on speaking proficiency. However, as illustrated in Figure 1, the MSE loss inevitably suffers from a limitation that different categories of phonemes that belong to the same proficiency level are inadvertently forced to be close to one another.

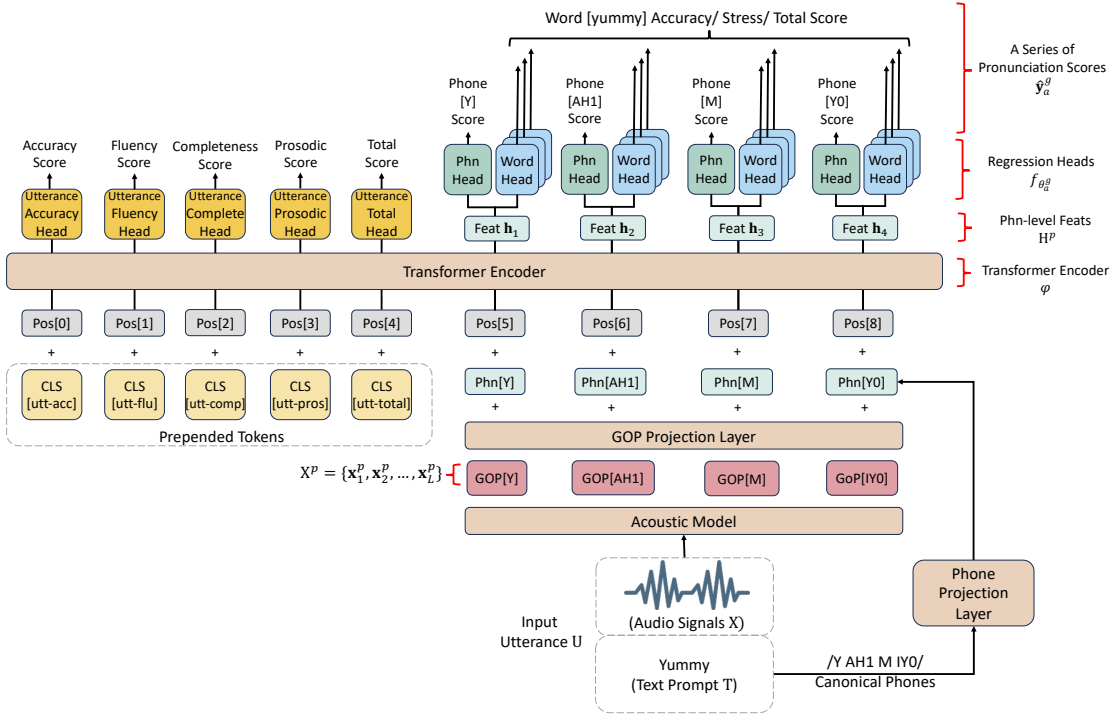

**Fig. 2**. A schematic illustration of the GOPT model [15].

To address this issue, we propose a novel loss function, dubbed phonemic contrast ordinal (PCO) loss, for enhancing regression-based APA models by introducing a phoneme-distinct regularizer into the MSE loss. This regularizer with two decoupled mathematical terms, namely the phonemic distinction and the ordinal tightness, manipulates the distances between inter- and intra-phoneme categories in both the feature and target spaces, respectively. The phonemic distinction concentrates on increasing the distances between feature centers of inter-phoneme categories, thereby ensuring that different phoneme categories can stay far apart from one another. The ordinal tightness preserves the ordinal relationships of the target output by considering the proficiency levels when pulling closer the intra-phoneme distances for each phoneme category. The synergy of these two terms provides the PCO loss with the capability to overcome the shortage of the MSE loss commonly adopted by existing APA models. To evaluate the effectiveness of the proposed PCO loss, we pair it with an iconic regression-based APA model, referred to as Goodness of Pronunciation feature-based Transformer (GOPT). GOPT adopts a transformer architecture as the backbone and simultaneously predicts proficiency scores at multiple linguistic granularities with various aspects [15].

## 2. PROPOSED METHOD

In this section, we begin by elaborating on the problem formulation of APA. Next, we give a brief sketch of the baseline APA model, GOPT. Finally, we introduce our proposed PCO loss for use in APA.

### 2.1. Problem Definition

We tackle the automatic pronunciation assessment (APA) task with the following processing flow. Given an input utterance U, which consists of a sequence of audio signals X uttered by an L2 learner, and a text prompt T that the learner is expected to pronounce it correctly. Our APA model manages to estimate a rich set of proficiency scores Y = $\{\mathbf{y}_a^g | g \in G, a \in A\}$ that covers different linguistic granularities G and aspects A. Here $\mathbf{y}_a^g$ stands for a vector of pronunciation scores, and its length is associated with the combination of linguistic granularity $g$ and aspect $a$. Specifically, for the input utterance U, the APA model trained to estimate five aspect scores at the utterance-level (i.e., accuracy, fluency, completeness, prosody and total score), qualify three aspect scores at the word-level (i.e., accuracy, stress and total score), and access an accuracy score at the phoneme-level. The aspect scores at utterance- and word-levels range from 0-10, while those of the phoneme-level aspects range from 0-2 in our experiments.

### 2.2. Revisiting the GOPT Model

GOPT, as shown in Figure 2, is a single, unified network composed of a transformer encoder network $\varphi$ and several aspect-specific regression heads $f_\Theta$, where $f_\Theta$ collectively is a regression function with a parameter set $\Theta = \{\theta_a^g | g \in G, a \in A\}$. For an input utterance U, the GOPT model first aligns the audio signals X with the text prompt T to extract a sequence of phoneme-level goodness of pronunciation (GOP) features $X^p = (\mathbf{x}_1^p, \mathbf{x}_2^p, ..., \mathbf{x}_L^p)$, where each GOP feature $\mathbf{x}_l^p$ is derived from a combination of log phone posterior (LPP) [17] and log posterior ratio (LPR) [20]. Next, the encoder $\varphi$ takes as input the GOP feature sequence $X^p$ to produce high-level representations $H = \varphi(X^p)$, where H includes additional utterance-level representations corresponding to five trainable aspect embeddings prepended to the embeddings of $X^p$. After that, regression heads $f_\Theta$ are built upon H to predict corresponding cross-level, aspect-specific proficiency scores in parallel, resulting in predicted scores $\widehat{Y} = \{\hat{\mathbf{y}}_a^g | g \in G, a \in A\}$, where $\hat{\mathbf{y}}_a^g = f_{\theta_a^g}(H)$. The encoder $\varphi$ and the regressor $\Theta$ are learned by minimizing the MSE loss:

$$\mathcal{L}_{mse} = \frac{1}{|G| \times |A|} \sum_{g \in G} \sum_{a \in A} \left\| \mathbf{y}_a^g - \hat{\mathbf{y}}_a^g \right\|_2^2, \quad (1)$$

where $\|.\|_2$ denotes the L2 norm.

### 2.3. Phonemic Contrast Ordinal Loss

As shown in Figure 1, the regression-based APA model like GOPT, trained using the MSE loss, can capture the ordinal relationship of the proficiency levels but tend to neglect the distinction between phoneme categories, leading to the aggregation of different categories of phonemes with the same proficiency level. As a remedy, we propose a phonemic contrast ordinal (PCO) loss by introducing a phoneme-distinct regularizer into mean-squared error loss. This regularizer consists of two mathematical terms, namely the phonemic distinction and the ordinal tightness, which seamlessly work together to regulate the distances within and between the phoneme categories in both feature and target space, respectively.

Applying the proposed loss function to different regression heads at various linguistic levels has the potential to improve the performance of the associated aspects. However, as an initial attempt, this paper focus on design a novel loss function that merely considers the phoneme-level representation features (i.e., intermediate embeddings) $H^p = (\mathbf{h}_1, \mathbf{h}_2, ..., \mathbf{h}_L)$ and their associated scores $\mathbf{y}^p = (y_1^p, y_2^p, ..., y_L^p)$, where $H^p$ simply exclude the [CLS] token from H.

**Phonemic Distinction.** The proposed PCO loss adopts a phonemic distinction term $\mathcal{L}_{pd}$ to address the distances between inter- phoneme categories in the feature space. Specifically, the phonemic distinction term $\mathcal{L}_{pd}$ aims to minimize negative distances between feature centers $\mathbf{h}_{c_i}$, which is equivalent to maximizing the distances between phoneme categories during the optimization process. The feature centers $\mathbf{h}_{c_i}$ are calculated by taking a mean over all some of the representation features $H^p$ that belong to the same phoneme category $p_i$. We define the phonemic distinction term $\mathcal{L}_{pd}$ by

$$\mathcal{L}_{pd} = -\frac{1}{M(M-1)} \sum_{i=1}^{M} \sum_{i \neq j} \left\| \mathbf{h}_{c_i} - \mathbf{h}_{c_j} \right\|_2 \times m_c, \quad (2)$$

where $M$ is the number of feature centers in a batch of samples or a sampled subset from a batch, and $m_c$ is a positive hyper-parameter that stands for a margin to the distance between two feature centers. Notably, owing to the unbounded nature of the feature spaces, it is necessary to normalize the features $H^p$ before further processing. Furthermore, the hyper-parameter $m_c$ empirically set to be 1 in our experiments.

**Ordinal Tightness.** Simply spreading the features may break the inherent ordinality of the regression target output. As a remedy, the ordinal tightness preserves the ordinal relationships by considering the proficiency level when reducing the scatterness of each phoneme category.

Formally, we introduce an additional ordinal tightness term $\mathcal{L}_{ot}$ that minimizes the distance between each representation feature $\mathbf{h}_i$ with its feature centers $\mathbf{h}_{c_i}$ while being aware of pronunciation score in the label (output) space:

$$\mathcal{L}_{ot} = \frac{1}{N} \sum_{i=1}^{N} \left\| \mathbf{h}_i - \mathbf{h}_{c_i} \right\|_2 \times y_i^p, \quad (3)$$

where $N$ is the batch size, and $\mathbf{y}_i$ is the corresponding phoneme-level accuracy score. Adding this ordinal tightness term encourages features with higher pronunciation scores to be closer to their centers while those with lower scores are farther away. The final formula of the PCO loss is defined by

$$\mathcal{L}_{pco} = \mathcal{L}_{mse} + \lambda_d \mathcal{L}_{pd} + \lambda_o \mathcal{L}_{ot}, \quad (4)$$

where $\mathcal{L}_{mse}$ is the MSE loss; and $\lambda_d$ and $\lambda_o$ are tunable parameters (set by cross-validation) that control the trade-off

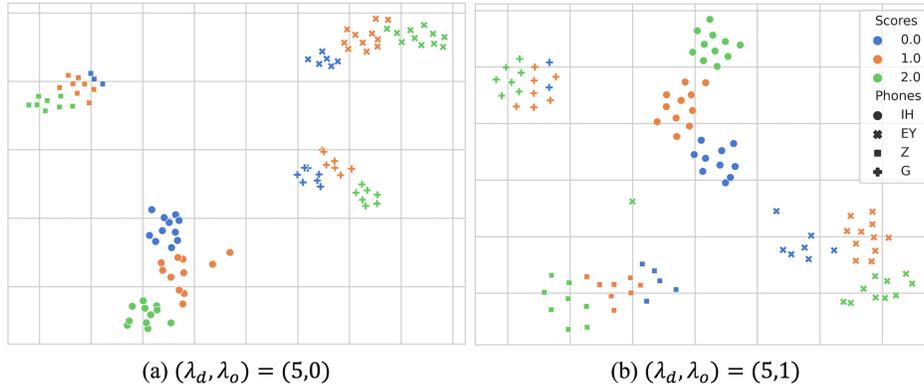

(a) $(\lambda_d, \lambda_o) = (5,0)$  (b) $(\lambda_d, \lambda_o) = (5,1)$

**Fig. 3**. Visualization of phoneme representations learned from APA (GOPT) models trained with proposed PCO loss, we show (a) the effect of phonemic distinction $\mathcal{L}_{pd}$ with hyperparameters $(\lambda_d, \lambda_o) = (5,0)$, and (b) the synergy of using both phonemic distinction and ordinal tightness $\mathcal{L}_{ot}$ with hyperparameters $(\lambda_d, \lambda_o) = (5,1)$.

balance between the phonemic distinction term $\mathcal{L}_{pd}$ and the ordinal tightness term $\mathcal{L}_{ot}$

## 3. EXPERIMENTS

### 3.1. Dataset

We conducted APA experiments on the speechocean762 dataset, which is a publicly available open-source dataset specifically designed for pronunciation assessment [21]. This dataset contains 5,000 English-speaking recordings spoken by 250 Mandarin L2 learners. The training and test sets are of equal size, each of which has 2,500 utterances. Speechocean762 contains comprehensive annotation information, where pronunciation proficiency scores were evaluated at multiple linguistic granularities with various aspects. Each score was independently assigned by five experts using the same rubrics, and the final score was determined by selecting the median value from the five scores. As with [15], we normalized utterance-level and word-level scores to the same scale as the phone score (0-2) for training APA models.

### 3.2. Implementation Details

Following settings presented in [15], we adopted the same DNN-HMM acoustic models to extract 84-dimensional GOP features. This acoustic model was based on a factorized time-delay neural network (TDNN-F) and trained using the Librispeech 960-hour data with the widely-used Kaldi recipe [22]. In order to evaluate the effectiveness of our proposed loss function, we kept all training hyper-parameters of GOPT compliant with the settings described in [15]. The number of transformer block was set to be 3, with 24 hidden units in each block. Each regression head is designed to use only one layer for projecting the hidden representation features to their corresponding pronunciation scores.

To ensure the reliability of our experimental results, we repeated 5 independent trials, each consisting of 100 epochs, using different random seeds. The experimental results are reported by averaging the top 100 best-performing experiments based on their Pearson Correlation Coefficient (PCC) performance on the training set. The primary evaluation metric is PCC, which measures the linear correlation between predicted scores and ground-truth scores. In addition, MSE value is used to assess phoneme-level accuracy.

## 4. EXPERIMENTAL RESULTS

### 4.1. Visualization of Phoneme Representations

Before launching into a series of experiments on the APA task, we visualize the phoneme-level intermediate embeddings of the vanilla GOPT models optimized with the proposed loss function under different settings, as shown in Figure 3. In so doing, we can graphically examine the effects of phonemic distinction and ordinal tightness on the phone-distinct regularizer. As demonstrated in Figure 3(a), it is evident that the PCO loss with the phonemic distinction term can effectively encourage feature representations to scatter apart according to their respective phoneme categories. We further plot the phoneme representations when adding the ordinal tightness term in the Figure 3(b). From this figure, we can observe that when the ordinal tightness term is added, the phoneme representations with a proficiency score of 0 in different phoneme categories are distributed further apart, especially in the phoneme category of /EY/, /IH/ and /Z/. Due to this property, GOPT trained with the proposed PCO loss can significantly enhance the discriminability of phone representations, meanwhile making them more sensitive to the corresponding proficiency scores.

**Table 1**. Comparisons of performance among various APA models on speechocean762.

| Model | Phone-level | | Word-level (PCC) | | | Utterance-level (PCC) | | | | |
|---|---|---|---|---|---|---|---|---|---|---|
| | MSE | PCC | ACC | Stress | Total | ACC | Completeness | Fluency | Prosody | Total |
| GOPT [15] | 0.085 | 0.612 | 0.533 | 0.291 | 0.549 | 0.714 | 0.155 | 0.753 | 0.760 | 0.742 |
| HiPAMA [17] | 0.084 | 0.616 | **0.575** | 0.320 | **0.591** | **0.730** | 0.276 | 0.749 | 0.751 | **0.754** |
| SB$_{num}$ Loss [18] | 0.086 | 0.605 | 0.531 | **0.386** | 0.547 | 0.722 | **0.427** | 0.750 | 0.752 | 0.747 |
| PCO Loss | **0.083** | **0.622** | 0.558 | 0.250 | 0.573 | 0.727 | 0.359 | **0.763** | **0.763** | 0.752 |

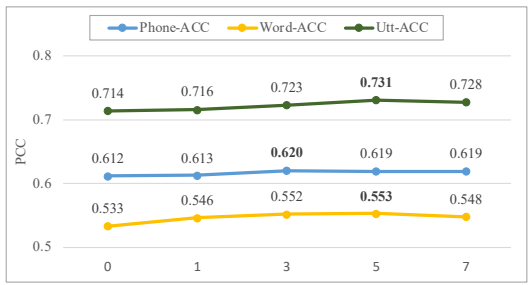

(a) PCC results *w.r.t.* parameter $\lambda_d$.

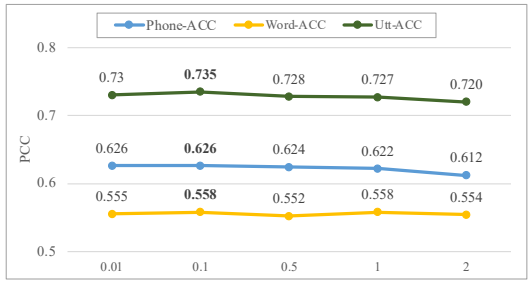

(b) PCC results *w.r.t.* parameter $\lambda_o$.

**Fig. 4**. Comparisons of the PCC results with respect to different settings of hyper-parameters $\lambda_d$ and $\lambda_o$.

### 4.2. Overall Performance on the APA Task

In the second set of experiments, we discuss the overall performance of our model in comparison with some other completive models. To better verify the effectiveness of our model on the APA task, we curate three top competitive models stemming from the GOPT model, which were also trained on speechocean762 without resort to any additional speech datasets. The corresponding results are illustrated in Table 1, from which we can make several observations. First, the proposed method (denoted by PCO Loss) consistently outperforms GOPT across three linguistic levels and various aspects. In addition, our method stands out in terms of the phoneme-level accuracy metric. Second, at the word-level granularity, our method consistently outperforms SBnum except for the stress evaluation. A possible reason for this is that the stress aspect in the word-level granularity suffers from the problem of severe data imbalance. As a side note SBnum proposed a loss reweighting scheme for the MSE loss so as to rebalance the loss contribution of frequent and infrequent prediction cases of different aspects independently, pertaining to their respectively training statistics [18]. Furthermore, our method slightly lags behind HiPAMA in the evaluations of the word-level granularity, which may be attributed to the better hierarchical model architecture of HiPAMA that works in conjunction with the aspect attention mechanisms. Third, in the utterance-level evaluation, our method not only appears to perform on par with HiPAMA in terms of the accuracy aspect and the total aspect, but also outperforms the other competitive models in terms of the fluency and prosody metrics which access the high-level pronunciation skills taking into account factors such as speaking style (e.g., repetition, stammering or hesitations), rhythm and intonation. Notably, our method can effectively distinguish phoneme representations belonging to lower-scoring groups from others by the ordinal tightness term, simultaneously separating the representations according to their phoneme categories with phonemic distinction term. These collectively bring benefits to the evaluations on the fluency and prosody aspects.

### 4.3. Ablation Studies

In the last set of experiments, we conducted ablation studies to analyze the contribution of each component involved in the PCO loss at three linguistic levels in terms of the PCC metric on the accuracy aspect.

**Effect of Phonemic Diversity.** We first analyze the effect of the phonemic diversity term by varying the parameter $\lambda_d$. Here we remove the ordinal tightness term (i.e., setting $\lambda_o = 0$), and report corresponding performance in Figure 4(a). From this figure, we can observe that the performance of the three linguistic levels is boosted as the value of $\lambda_d$ increases, but tends to reach a plateau when $\lambda_d$ exceeds 5. Furthermore, the best performance for the word- and utterance-level speaking proficiency evaluations is achieved when $\lambda_d$ is set to 5.

**Effect of Ordinal Tightness.** Next, we investigate the effectiveness of the ordinal tightness term by altering the value of parameter $\lambda_o$ while keeping $\lambda_d$ fixed at 5. As shown in Figure 4(b), we can observe a decreasing trend in the accuracy evaluations of all granularity levels as the value of $\lambda_o$ becomes larger. When $\lambda_o$ is set equal to 0.1, the associated evaluations of all granularity level yield the best performance.

## 5. CONCULSION AND FUTURE WORK

This paper has put forward a simple yet effective loss function, dubbed the phonemic contrast ordinal (PCO) loss for the APA task. The PCO loss introduces a phoneme-distinct regularizer into the MSE loss to regulate the distances between inter- and intra-phoneme categories in both the feature and target spaces. A series of empirical experiments conducted on the speechocean762 benchmark dataset has revealed the feasibility of our proposed model in comparison to some top-of-the-line models. In future work, we plan to pair the PCO loss with more sophisticated model structures that can integrate lexical and phonological cues, as well as context-aware hierarchical information [24][25][26].

## 6. ACKNOWLEDGEMENTS

We are grateful to all the anonymous reviewers for their helpful advice on various aspects of this work. This work was supported in part by E.SUN bank under Grant No. 202308-NTU-03. Any findings and implications in the paper do not necessarily reflect those of the sponsors.